\documentclass[aps,prd,twocolumn,showpacs,preprintnumbers,amsmath,amssymb]{revtex4}
\usepackage{amsfonts}

\usepackage{graphicx}
\usepackage{dcolumn}
\usepackage{bm}

\begin{document}
\title{The bound state problem of S-wave heavy quark meson-aitimeson systems}

\author{Yan-Rui Liu}
\email{yrliu@ihep.ac.cn}

\author{Zong-Ye Zhang}
\email{zhangzy@ihep.ac.cn}

\affiliation{1. Institute of High Energy Physics, CAS, P.O. Box
918-4, Beijing 100049, China\\
2. Theoretical Physics Center for Science Facilities, CAS, Beijing
100049, China}

\date{\today}

\begin{abstract}

We investigated systematically whether the S-wave ($\bar{Q}q$) meson
and the ($Q\bar{q}$) meson may form S-wave bound states in a chiral
SU(3) quark model by solving the resonating group method equation.
Here $Q=c$ or $b$ and $q=u$, $d$ or $s$. Our preliminary calculation
disfavors the existence of $I=\frac12$ ($\bar{Q}l$)-($Q\bar{s}$)
molecules ($l=u,d$) while favors the existence of isoscalar
$B\bar{B}$, $B^*\bar{B}^*$ (J=2) and $B\bar{B}^*$ (C=+) molecules.
The existence of isovector (charm-anticharm) and (charm-bottom)
molecules is also disfavored. Therefore the resonance-like structure
$Z^+(4051)$ is unlikely to be an S-wave $D^*\bar{D}^*$ molecule.

\end{abstract}


\maketitle

\section{introduction}\label{sec1}

The molecular picture was widely used in discussing the strange
states, such as $f_0(980), a_0(980)$
\cite{uni-980,wein-980,pot,eft,inf,bs-980} and $\Lambda(1405)$
\cite{lambda,ini,uni-140}. Although it is still difficult to
identify an exotic state as a hadronic molecule, the exploration for
possible molecules in more systems is an interesting topic. Such a
study may help us to understand the strong interactions. There were
dynamical studies whether the possible molecules exist in the light
quark systems. In Refs. \cite{KN-c,Kstrange,omemeson,Nphi}, various
meson-baryon systems were investigated, while in Refs.
\cite{OmegaNbar,NNbar,nnbar-cpc,nnbar-had,1835-baryonimum,1835-fsi,NNbar-model},
the $\bar{\Omega} N$ and $N\bar{N}$ systems were studied.

For heavy quark systems, the formation of molecules is easier due to
the relatively small kinetic term. The relevant study can be traced
back to thirty years ago \cite{Okun,RGG}. Ten years later,
T\"ornqvist studied possible deuteron-like meson-meson states bound
by pions in Refs. \cite{Tornqvist-p,Tornqvist-s} which were called
deusons \cite{Tornqvist-t}. In Ref. \cite{cri-mas}, Ericson and Karl
investigated the critical mass for molecule formation. In recent
years, the renaissance of hadron spectroscopy, especially the
observation of exotic heavy quark mesons
\cite{Hmesons,charmonium,XYZ,newHadron}, triggered extensive
discussions in the molecular picture.

The charmed meson $D_{sJ}(2317)$ \cite{2317-ex,2460-ex} whose mass
is much smaller than the quark model prediction was once proposed as
a $DK$ molecule \cite{barnes-2317}. Similarly, $D_{sJ}(2460)$
\cite{2460-ex} was suggested as a $D^*K$ state. However, their
$c\bar{s}$ nature is strongly favored after considering the
significant contributions from the $DK$ continuum \cite{zhu-2317}.

The discovery of X(3872)
\cite{3872-first,3872-CDF,3872-D0,3872-BaBar} ignited physicists'
great interests. It is almost on the threshold of $D^0\bar{D}^{*0}$
and very close to the thresholds of $\rho J/\psi$, $\omega J/\psi$
and $D^+D^{\ast-}$. The most popular interpretation for this
intriguing state is a hadronic molecule dominated with
$D^0\bar{D}^{*0}$
\cite{Mole-Close,Mole-Voloshin,Mole-Torqvist,Mole-Swanson,Mole-Wong}.
However, this picture was questioned in Ref. \cite{suzuki}. Very
recently, the BaBar collaboration measured a relatively large
branching fraction for $X(3872)\to \psi(2S)\gamma$, which indicates
that $X(3872)$ is possibly a mixing state of $c\bar{c}$ and
$D^0\bar{D}^{*0}$ \cite{mixingX-exp}.

For the interpretations of the exotic Y(4260)
\cite{4260-babar,4260-cleo,4260-belle} in the molecular picture, Liu
et al. suggested it is a $\chi_{c1}\rho$ state \cite{xiang-4260}
while Yuan et al. proposed it is a $\chi_{c1}\omega$ state
\cite{yuan-4260}. There are also other molecular proposals such as a
$\Lambda_c$ pair \cite{qiao-4260} and a $D_0\bar{D}^*$ or
$D_1\bar{D}$ bound state \cite{nielsen-4260,ding-4260}. In fact, the
most popular opinion is that Y(4260) is a hybrid state
\cite{zhu-4260,close-4260,kou-4260} although this interpretation is
also inconclusive \cite{felipe-4260,chiu-4260}. We still require
detailed investigations to answer whether these interpretations are
correct or not.

Recently, the Belle collaboration observed a charged charmonium-like
state $Z^+(4430)$ in the $\pi^+\psi^\prime$ invariant mass
distribution \cite{Z4430}. This state is an excellent candidate of
heavy quark molecules. The dynamical calculation also indicates
$Z^+(4430)$ may be interpreted as a $D_1D^*$ ($D_1^\prime D^*$)
molecule
\cite{rosner-4430,chao-4430,liu-4430,ding-4430,suhoung-4430}. Not
long ago, the Belle collaboration announced two more charged
charmonium-like resonances $Z^+(4051)$ and $Z^+(4248)$ in the
$\pi^+\chi_{c1}$ mass distribution \cite{twocharged}, which gives us
the hope that heavy quark molecules do exist. Unfortunately, the
BaBar data do not support the existence of $Z^+(4430)$
\cite{4430-babar}. Cross-checks for the other two charged resonances
are also desired.

Therefore, none of the heavy quark molecules has been established
yet. With the development of experimental measurements, more and
more exotic states in the heavy quark region will be found. It is
worthwhile to study in which systems heavy molecules can exist.
Motivated by the observation of new exotic states and the
possibility of forming heavy quark molecules, Wong explored the
combinations of heavy mesons and heavy antimesons in a quark-based
model and found many molecular states \cite{Mole-Wong}. Voloshin and
Dubynskiy suggested the possible resonance at the $D^*\bar{D}^*$
threshold in Refs. \cite{Vec-DD1,Vec-DD2}. Zhang et al. studied
possible S-wave bound states of two pseudoscalar mesons using the
vector-meson-exchange potential \cite{zhangyj}. In Ref.
\cite{lee-mole}, a $D_sD^*$ molecule was proposed.

In a previous work \cite{liu-sys}, we studied the S-wave
$D\bar{D}/B\bar{B}$ $D^*\bar{D}^*/B^*\bar{B}^*$ and
$D^*\bar{D}/B^*\bar{B}$ systems in a meson exchange model at hadron
level, where we considered scalar, pseudoscalar and vector mesons
exchanges. In this paper, we will explore similar systems in a
chiral SU(3) quark model ($\chi$QM) \cite{SU3CQM} and calculate the
binding energies by solving the resonating group method (RGM)
equation \cite{Oka}. All the mesons below 1.1 GeV will be
considered. The study can be used to test different model
approaches.

The chiral quark model is a useful tool in connecting the QCD theory
and the experimental observables. It has been proved successful in
studying the baryon-baryon interactions and the meson-baryon
interactions. For the mechanism of the short range quark-quark
interaction, it is still controversial whether one-gluon exchange
(OGE) or vector-meson exchange dominates. In Ref. \cite{ExCQM}, Dai
et al. extended the chiral SU(3) quark model to include the vector
meson exchange part and named the model the extended chiral SU(3)
quark model (E$\chi$QM) which was also successful in reproducing the
energies of the baryon states, the binding energy of the deuteron
and the NN scattering phase shifts.

It is interesting to study whether this phenomenological approach is
applicable to the heavy quark systems. We have applied this model to
the $D^0\bar{D}^{*0}$ system in Ref. \cite{3872-rgm} and we will
continue to perform similar studies to other systems. One may test
this model by comparing the predictions with future measurements.

The paper is organized as follows. After the introduction, we
present a brief discussion about the systems we will study in
Section \ref{sec2}. In Section \ref{sec3}, we present the
ingredients of the model. Then in Section \ref{sec4}, we give the
essential parameters for the calculation. We show numerical results
for different systems in Section \ref{sec5}. The last section is the
discussion and summary.

\section{Heavy quark meson-antimeson systems}\label{sec2}

The S-wave single heavy quark mesons are pseudoscalar type [$D$,
$D_s$, $B$, $B_s$] and vector type [$D^*$, $D_s^*$, $B^*$, $B_s^*$].
For simplicity, $P$ ($V$) will represent the heavy quark
pseudoscalar (vector) meson. We investigate whether the hadronic
molecules can be found in the combinations of these mesons and their
antiparticles in this paper. From the flavor SU(3) symmetry, the
multiplets are $3\times \bar{3}=8+1$. One may consult Ref.
\cite{liu-sys} for the explicit flavor wave functions. The largely
broken SU(3) symmetry must be taken into account for possible
hadronic molecules. In the numerical evaluation, we will first
consider the isospin symmetric case. Because isospin symmetry
breaking (ISB) is probably important, we will also discuss the case
of large isospin breaking.

In the isospin symmetric case, we need consider only four
possibilities: (1). I=1/2 ($\bar{Q}u$)-($Q\bar{s}$), (2). I=1
($\bar{Q}u$)-($Q\bar{d}$), (3). I=0 ($\bar{Q}s$)-($Q\bar{s}$) and
(4). I=0 ($\bar{Q}l$)-($Q\bar{l}$), where $Q$ is a charm or bottom
quark and $l$ represents an up or down quark. We call them I=1/2,
I=1, I=0($s$) and I=0($l$) states respectively in the following
parts.

When studying the possible heavy molecule composed of a heavy meson
and an antimeson, we take a simple picture where only color-singlet
mesons are involved. The OGE and the confinement interactions occur
inside the mesons, while the meson exchange interaction occurs
between light quarks of different mesons. To make the description
accurate, we label the heavy quarks 1 and 3 while the light quarks 2
and 4. The quarks 1 and 2 are bound to the meson and the quarks 3
and 4 to the antimeson. We do not consider the flavor-singlet meson
exchange between heavy quarks or between a heavy quark and a light
quark in present investigation. The consideration is as follows. In
the chiral quark model, the constituent mass of the light quark is
related to the spontaneous vacuum breaking while the breaking gives
small effects to the masses of the heavy quarks, which indicates the
coupling of the sigma meson and the heavy quarks is weak.

\section{Hamiltonian}\label{sec3}

The details of the chiral SU(3) quark model can be found in Refs.
\cite{SU3CQM,ExCQM}. Here we just present essential constituents for
the calculation. The Hamiltonian for the meson-antimeson system has
the form
\begin{eqnarray}\label{ham}
H &=&\sum_{i=1}^4 T_i -T_G + V^{OGE}+V^{conf}+\sum_{M}V^M
\end{eqnarray}
where $T_i$ is the kinetic term of the $i$th quark or antiquark and
$T_G$ is the kinetic energy operator of the center of mass motion.
$M$ is the exchanged meson between light quarks.

The potential of the OGE part reads
\begin{eqnarray}
V_{\bar{q}Q}^{OGE}&=&g_qg_Q
\mathbf{F}^c_{\bar{q}}\cdot\mathbf{F}^c_Q\left\{\frac{1}{r}-\frac{\pi}{2}\delta^3(\mathbf{r})\Big[\frac{1}{m_q^2}+\frac{1}{m_Q^2}\right.\nonumber\\
&&\left.+\frac43\frac{1}{m_qm_Q}(\bm{\sigma}_q\cdot\bm{\sigma}_Q)\Big]\right\},
\end{eqnarray}
where $\mathbf{F}^c_{Q}=\frac{\bm{\lambda}}{2}$ for quarks and
$\mathbf{F}^c_{\bar{q}}=-\frac{\bm{\lambda}^\ast}{2}$ for
antiquarks. $m_q$ $(m_Q)$ is the light (heavy) quark mass. The
linear confinement potential is
\begin{eqnarray*}
V_{\bar{q}Q}^{conf}=-4\mathbf{F}^c_{\bar{q}}\cdot\mathbf{F}^c_Q\left(a^c_{qQ}
r +a^{c0}_{qQ}\right).
\end{eqnarray*}
There are similar expressions for $V_{q\bar{Q}}^{OGE}$ and
$V_{q\bar{Q}}^{conf}$.

For a molecule formed with $(Q\bar{q})$ and $(\bar{Q}q)$ mesons, the
light meson exchange occurs only between $\bar{q}$ and $q$. From
Refs. \cite{SU3CQM,ExCQM}, one gets
\begin{eqnarray}
V^{\sigma_a}(\bm{r}_{ij})&=&-C(g_{ch},m_{\sigma_a},\Lambda)X_1(m_{\sigma_a},\Lambda,r_{ij})[\lambda_a(i)\lambda_a(j)],\nonumber\\
&&(a=0,1,2, \cdots, 8)\\
V^{\pi_a}(\bm{r}_{ij})&=&C(g_{ch},m_{\pi_a},\Lambda)\frac{m_{\pi_a}^2}{12
m_2m_4}
X_2(m_{\pi_a},\Lambda,r_{ij})\nonumber\\
&&\times[\bm{\sigma}(i)\cdot\bm{\sigma}(j)][\lambda_a(i)\lambda_a(j)],\\
V^{\rho_a}(\bm{r}_{ij})&=&C(g_{chv},m_{\rho_a},\Lambda)\Big\{X_1(m_{\rho_a},\Lambda,r_{ij})
+\frac{m_{{\rho_a}}^2}{6
m_2m_4}\nonumber\\
&&\times\left[1+\frac{f_{chv}}{g_{chv}}\frac{m_2+m_4}{M_N}+
(\frac{f_{chv}}{g_{chv}})^2\frac{m_2m_4}{M_N^2}\right]\nonumber\\
&&\times
X_2(m_{{\rho_a}},\Lambda,r_{ij})[\bm{\sigma}(i)\cdot\bm{\sigma}(j)]\Big\}[\lambda_a(i)\lambda_a(j)],\nonumber\\\\
V_{q\bar{q}}^M&=&G_M V_{qq}^M.
\end{eqnarray}
Where $G_M$ is the G-parity of the exchanged meson and
\begin{eqnarray}
C(g_{ch},m,\Lambda)&=&\frac{g_{ch}^2}{4\pi}\frac{\Lambda^2
m}{\Lambda^2-m^2},\\
X_1(m,\Lambda,r)&=&Y(mr)-\frac{\Lambda}{m}Y(\Lambda r),\\
X_2(m,\Lambda,r)&=&Y(mr)-\left(\frac{\Lambda}{m}\right)^3 Y(\Lambda r),\\
Y(x)&=&\frac{e^{-x}}{x}.
\end{eqnarray}

Here we do not present the tensor term and the spin-orbital term in
the potentials since we consider only S-wave interactions. We use
the same cutoff $\Lambda$ for various mesons. Its value is around
the scale of chiral symmetry breaking ($\sim$1 GeV).

By solving the RGM equation, one gets the energy of the system and
the relative motion wave function. From the definition of the
binding energy, $E_0=M_{\bar{Q}q}+M_{Q\bar{q}}-M_{system}$, one
judges whether a system would be bound or not.

\section{The parameters}\label{sec4}

There are several parameters in the Hamiltonian and the wave
functions: the OGE coupling constants $g_q$ and $g_Q$, the
confinement strengths $a_{qQ}^c$, the zero-point energies
$a_{qQ}^{c0}$, the quark masses $m_Q$ and $m_q$, the
harmonic-oscillator width parameter $b_u$, the quark-meson coupling
constants $g_{ch}$, $g_{chv}$ and $f_{chv}$, the cutoff $\Lambda$
and the mixing angle for the $I=0$ mesons. The mass of the
phenomenological $\sigma$ meson is also treated as an adjustable
parameter. For other meson masses, we use the experimental values.

The sigma meson does not have a definite mass. In the light quark
systems, this mass parameter was adjusted to fit the mass of the
baryons, the binding energy of the deuteron and the NN phase shifts.
When extending the application of this model to the heavy quark
systems, we use the values determined in the light quark systems. If
the vector meson exchanges are not included, the mass is 595 MeV,
while $m_\sigma=535$ MeV and 547 MeV were used in the E$\chi$QM.

For the up and strange quark masses, we use the values given in the
previous work \cite{SU3CQM,ExCQM,OmegaNbar}, $m_u=313$ MeV and
$m_s=470$ MeV. To investigate the heavy quark mass dependence, we
take several typical values $m_c=1430$ MeV \cite{zzz}, $m_c=1870$
MeV \cite{Semay}, $m_b=4720$ MeV which is close to the value in Ref.
\cite{zzz2}, and $m_b$=5259 MeV \cite{Semay}.

The chiral coupling constant $g_{ch}$ is related to $g_{NN\pi}$
through
\begin{eqnarray}
\frac{g_{ch}^2}{4\pi}=\frac{9}{25}\frac{g_{NN\pi}^2}{4\pi}\frac{m_u^2}{m_N^2}
\end{eqnarray}
with $g_{NN\pi}^2/(4\pi)=13.67$ determined experimentally. From this
relation, one gets $g_{ch}=2.621$. In the extended SU(3) chiral
quark model, one also needs the vector coupling constants. We adopt
two sets of the values used in the previous work,
$(g_{chv},f_{chv})$=(2.351,0.0), and (1.972,1.315) \cite{ExCQM}. The
corresponding sigma mass is also presented in Table \ref{paras}. One
notes each set of parameters can reproduce the masses of the ground
state baryons, the binding energy of the deuteron and the $NN$ and
$YN$ scattering observables.

\begin{table}
\centering
\begin{tabular}{ccccc}
\hline
    & $\chi$QM & \multicolumn{2}{c}{E$\chi$QM } \\
    &        Set 1      & Set 2  & Set 3 \\\hline
$b_u$ (fm)& 0.5 & 0.45 & 0.45\\
$m_u$ (MeV)& 313 & 313 & 313\\
$m_s$ (MeV)& 470 & 470 & 470\\
$m_\sigma$ (MeV)& 595 & 535 & 547\\
$g_{chv}$ &     & 2.351 & 1.972 \\
$f_{chv}/g_{chv}$ & & 0 & 2/3
\\\hline
\end{tabular}
\caption{Three sets of model parameters. Other meson masses are:
$m_{\sigma^\prime}=984.7$ MeV, $m_\epsilon=980$ MeV, $m_\pi=138$
MeV, $m_\eta=547.8$ MeV, $m_{\eta^\prime}=957.8$ MeV, $m_\rho=775.8$
MeV, $m_\omega=782.6$ MeV and $m_\phi=1020$ MeV.}\label{paras}
\end{table}

The values of $g_q$, $g_Q$, $a_{qQ}^c$ and $a_{qQ}^{c0}$ can be
derived from the masses of the ground state baryons and the heavy
mesons. The binding energy for a system of two color-singlet mesons
is irrelevant to the internal potentials of the color-singlet meson
because of the cancellation \cite{3872-rgm}. Therefore these four
values will not give effects to the final results of $E_0$. We do
not present them here.

Isoscalar states with the same $J^{PC}$ will mix. The mixing angle
for pseudoscalar mesons $\eta_1$ and $\eta_8$, $\theta^{PS}$, is
taken to be $-23^\circ$. Because the mixing angle $\theta^S$ for
scalar mesons is still unclear and controversial, we use three
values in the numerical evaluation: $0.0$, $35.264^\circ$ and
$-18^\circ$. The second number corresponds to the ideal mixing while
the last one is taken from Ref. \cite{dai-angle}. We use the ideal
mixing angle $\theta^V=35.264^\circ$ for the vector mesons. In the
scalar and pseudoscalar meson exchange potentials, we have adopted
$\lambda_0=\mathbb{I}$ where $\mathbb{I}$ is the unit matrix. To
investigate its effects, we also use
$\lambda_0=\sqrt{\frac23}\mathbb{I}$ to calculate the binding
energies.

To consider the dependence of the binding energy on the cutoff, we
use two values $\Lambda=1100$ MeV and $\Lambda=1500$ MeV.

\section{Numerical Results}
\label{sec5}

When performing the numerical evaluations, we calculate the binding
energies with all possible combinations of the parameters presented
in the previous section. Only when all the results for a system
indicate it is unbound, we conclude the system is unbound. On the
contrary, we say a molecule is possible only when all the results
indicate the system is bound.

\subsection{$P\bar{P}$ systems}

The quantum numbers for the neutral states are $J^{PC}=0^{++}$. The
pseudoscalar mesons do not exchange in such systems since the
coupling of three pseudoscalar mesons is forbidden.

For I=1/2 states, we investigate $\bar{D^0}D_s^+$, $B^+\bar{B}_s^0$
and $B^+D_s^+$. Such systems are possibly bound by only scalar
mesons $\sigma$ and $\epsilon$. After solving the RGM equation, we
find these systems are unbound with various parameters presented in
the previous section.

\begin{table}[htb]
\centering
\begin{tabular}{cc|cc}\hline
Isospin&System&{$\chi$QM} &{E$\chi$QM}\\
$I=1$&$\bar{D}^{0}D^{+}$ & $\times$ & $\times$   \\
&$B^{+}\bar{B}^{0}$& $\times$ & $\leq 1.9$ \\
&$B^{+}D^{+}$ & $\times$ & $\times$
\\ \hline
$I=0(s)$&$D_s^-D_s^+$ & $\times$ & $\leq10.4$   \\
&$B_s^0\bar{B}_s^0$& $\leq13.3$ & $2.5\sim43.7$ \\
&$B_s^0 D_s^+$ & $\leq3.0$ & $\leq22.6$
\\ \hline
$I=0(l)$&$D^-D^++\bar{D}^0D^0$ & $\leq4.9$ & $13.7\sim52.9$   \\
&$B^0\bar{B}^0+B^+B^-$& $10.1\sim26.8$ & $47.2\sim102.3$ \\
&$B^0D^++B^+D^0$ & $0.4\sim12.8$ & $23.1\sim72.8$
\\ \hline
ISB&$\bar{D}^0D^0$ & $\times$ & $\leq12.8$   \\
&$B^+B^-$& $0.1\sim10.3$ & $11.7\sim43.3$ \\
&$B^+D^0$ & $\leq2.0$ & $0.5\sim24.2$
\\ \hline
\end{tabular}
\caption{The binding energies for $P\bar{P}$ states. $\times$ means
the system is unbound.}\label{PPbar}
\end{table}

For I=1 systems, we calculate the binding energies of
$\bar{D}^0D^+$, $B^+\bar{B}^0$ and $B^+D^+$. Vector mesons $\rho$
and $\omega$ are permitted, but $\rho$ exchange interaction is
repulsive while $\omega$ is attractive. Their contributions are
almost canceled. $\sigma$ and $\epsilon$ provide attractive force
while $\sigma^\prime$ gives repulsive interaction. By exploring
different cases of parameters, we get the binding energies for these
systems. The final results are given in Table \ref{PPbar}. If all
the numerical values indicate the system is unbound, we mark it with
``$\times$''. If the system is unbound with some parameters and
bound with other parameters, we give the upper limit of the binding
energy. If all the results indicate the system is bound, we collect
the binding energies in a range. From the table, one knows there are
no bound states in $\bar{D}^0D^+$ and $B^+D^+$.

The hidden strange I=0($s$) states we investigate include
$D_s^-D_s^+$, $B_s^0\bar{B}_s^0$ and $B_s^0 D_s^+$. They may be
bound mainly by the attractive $\sigma$, $\epsilon$ and $\phi$.
According to our model calculation, it is difficult to draw a
definite conclusion whether the bound states may form (see Table
\ref{PPbar}).

The I=0($l$) systems we study are
$\frac{1}{\sqrt2}(D^-D^++\bar{D}^0D^0)$,
$\frac{1}{\sqrt2}(B^0\bar{B}^0+B^+B^-)$ and
$\frac{1}{\sqrt2}(B^0D^++B^+D^0)$. Comparing with the I=1 systems,
$\sigma^\prime$ and $\rho$ exchange interactions are both attractive
now. The amplitudes of the potentials are also larger. From the
results in Table \ref{PPbar}, we find the bound states containing
bottom quarks exist, even if only scalar mesons can exchange.

In the real world, the isospin symmetry is also broken. The mass
difference between $D^0$ and $D^\pm$ is around 5 MeV and it will
affect the conclusion whether hadronic molecules exist or not. In
this study, we also calculate preliminarily the extreme cases
$\bar{D}^0D^0$, $B^+B^-$ and $B^+D^0$. Such cases get the minimum
contributions from $\sigma^\prime$ and $\rho$. Our results indicate
the hidden bottom molecule $B\bar{B}$ is still possible. Table
\ref{PPbar} shows relevant results.

\subsection{$V\bar{V}$ systems}

The quantum numbers are $J^{PC}=0^{++}$, $1^{+-}$, or $2^{++}$ for
the neutral states. The pseudoscalar mesons, scalar mesons and
vector mesons can all be exchanged in such systems. In our model,
the amplitudes for scalar meson exchange interactions are the same
as the $P\bar{P}$ case.

Similar to the former case, we first investigate the
$\bar{D}^{*0}D_s^{*+}$, $B^{*+}\bar{B}_s^{*0}$ and $B^{*+}D_s^{*+}$
systems with I=1/2. Here the vector meson exchanges are forbidden.
The contributions from $\eta$ and $\eta^\prime$ cancel largely and
the pseudoscalar mesons give finally small contributions. The
$\sigma$ and $\epsilon$ have not enough attractive force to bind the
heavy mesons and these systems are unbound for the angular momentum
$J$=0, 1, and 2.

We explore three I=1 systems $\bar{D}^{*0}D^{*+}$,
$B^{*+}\bar{B}^{*0}$ and $B^{*+}D^{*+}$. Comparing with I=1
$P\bar{P}$ case, the exchanges of pseudoscalar mesons $\pi$, $\eta$
and $\eta^\prime$ are permitted. The contributions from $\eta$ and
$\eta^\prime$ reduce that from $\pi$. For $J=0$ and $J=1$, the
interaction due to pseudoscalar mesons is attractive and for $J=2$,
it is repulsive. From the resulting binding energies, we conclude
that $\bar{D}^{*0}D^{*+}$ and $B^{*+}D^{*+}$ are not bound while
$B^{*+}\bar{B}^{*0}$ is not excluded. We present our results in
Table \ref{VVbar}.

The hidden strange states (I=0) include $D_s^{*-}D_s^{*+}$,
$B_s^{*0}\bar{B}_s^{*0}$ and $B_s^{*0} D_s^{*+}$. The contributions
from $\eta$ and $\eta^\prime$ exchange interactions have the same
sign. For $J=0$ and $J=1$, they are repulsive. For $J=2$, they are
attractive. Our numerical results are also presented in Table
\ref{VVbar}. $D_s^{*-}D_s^{*+}$ is not bound in $\chi$QM.

For I=0($l$) systems, we calculate the binding energies of
$\frac{1}{\sqrt2}(D^{*-}D^{*+}+\bar{D}^{*0}D^{*0})$,
$\frac{1}{\sqrt2}(B^{*0}\bar{B}^{*0}+B^{*+}B^{*-})$ and
$\frac{1}{\sqrt2}(B^{*0}D^{*+}+B^{*+}D^{*0})$. The $\pi$, $\eta$ and
$\eta^\prime$ exchange interactions have like sign. For $J=0$ and
$J=1$, they are repulsive while they are attractive for $J=2$. We
find there are no binding solutions for these systems in $\chi$QM if
$J=0$ while the formation of molecules is possible if $J=2$. Table
\ref{VVbar} shows our results.

Similar to the $P\bar{P}$ isospin breaking case, we study whether
$\bar{D}^{*0}D^{*0}$, $B^{*+}B^{*-}$ and $B^{*+}D^{*0}$ may be
bound.  According to our calculation, bound states in $\chi$QM do
not exist if $J=0$ and the hidden bottom molecule is still possible
if $J=2$. We also present the results for this extreme case in Table
\ref{VVbar}.

\begin{table*}[t]
\centering
\begin{tabular}{cc|ccc|ccc}\hline
Isospin&System&\multicolumn{3}{c}{$\chi$QM} &\multicolumn{3}{c}{E$\chi$QM}\\
&& {$J=0$}& {$J=1$} & {$J=2$} & {$J=0$}& {$J=1$} & {$J=2$} \\
$I=1$&$\bar{D}^{*0}D^{*+}$ & $\times$ & $\times$ & $\times$ & $\times$ & $\times$ & $\times$ \\
&$B^{*+}\bar{B}^{*0}$& $\leq 3.5$ & $\leq 1.2$ & $\times$ & $\leq7.9$ & $\leq4.7$ & $\times$ \\
&$B^{*+}D^{*+}$ & $\times$ & $\times$ & $\times$ & $\times$ &
$\times$ & $\times$
\\ \hline
$I=0(s)$&$D_s^{*-}D_s^{*+}$ & $\times$ & $\times$ & $\times$ & $\leq 10.2$ & $\leq 10.3$ & $\leq 10.4$ \\
&$B_s^{*0}\bar{B}_s^{*0}$& $\leq 10.4$ & $\leq 11.8$ & $\leq 15.0$ & $3.3\sim43.7$ & $2.9\sim43.7$ & $2.1\sim43.7$ \\
&$B_s^{*0} D_s^{*+}$ & $\leq 1.8$ & $\leq 2.3$ & $\leq 3.8$ & $\leq
22.5$ & $\leq 22.6$ & $\leq 22.6$
\\ \hline
$I=0(l)$&$D^{*-}D^{*+}+\bar{D}^{*0}D^{*0}$ & $\times$ & $\times$ & $3.6\sim 22.4$ & $\leq 25.6$ & $3.5\sim38.7$ & $17.7\sim67.9$ \\
&$B^{*0}\bar{B}^{*0}+B^{*+}B^{*-}$& $\times$ & $\leq 5.8$ & $34.5\sim59.2$ & $22.6\sim66.4$ & $34.7\sim84.2$ & $60.2\sim120.8$ \\
&$B^{*0}D^{*+}+B^{*+}D^{*0}$ & $\times$ & $\leq 0.4$ &
$13.9\sim36.7$ & $5.6\sim41.4$ & $13.6\sim56.7$ & $33.4\sim89.5$
\\ \hline
ISB&$\bar{D}^{*0}D^{*0}$ & $\times$ & $\times$ & $\leq1.1$ & $\leq 6.7$ & $\leq9.6$ & $\leq16.2$ \\
&$B^{*+}B^{*-}$& $\times$ & $\leq 3.5$ & $5.3\sim20.2$ & $5.6\sim32.7$ & $8.6\sim37.9$ & $14.9\sim48.7$ \\
&$B^{*+}D^{*0}$ & $\times$ & $\times$ & $\leq7.5$ & $\leq16.0$ &
$\leq20.0$ & $2.3\sim28.6$
\\ \hline
\end{tabular}
\caption{The binding energies for $V\bar{V}$ states. $\times$ means
the system is unbound.}\label{VVbar}
\end{table*}

\subsection{$P\bar{V}\pm V\bar{P}$ systems}

The components $P\bar{V}$ and $V\bar{P}$ do not have definite
C-parity while the neutral $P\bar{V}\pm V\bar{P}$ states do. For a
state with given C-parity, two conventions for the relative sign
have been used in the literature. The plus sign for the C=+
$D\bar{D}^*$ system corresponding to the X(3872) was widely used
while the minus sign was adopted in Refs. \cite{liu-3872,Dong}.
Recently, Stancu analyzed the charge conjugation in multiquark
systems in detail \cite{stancu} and she also obtained a minus sign.
In fact, the convention of the relative sign depends on the phase
between $P$ and $\bar{P}$ as well as $V$ and $\bar{V}$ under the
charge conjugation transformation. But the final result is
irrelevant with the convention. For example, for the C=+
$D^0\bar{D}^{*0}$ state, one gets
$X=\frac{1}{\sqrt2}(D^0\bar{D}^{*0}-D^{*0}\bar{D}^0)$ with the
convention $D^0(D^{*0})=c\bar{u}$ and
$\bar{D}^0(\bar{D}^{*0})=u\bar{c}$. The resulting matrix element
$\langle X|{\bf \sigma_2}\cdot{\bf\sigma_4}|X\rangle$ is +1. If the
convention $D^0(D^{*0})=c\bar{u}$ and
$\bar{D}^0(\bar{D}^{*0})=\bar{c}u$ is used, one gets
X=$\frac{1}{\sqrt2}(D^0\bar{D}^{*0}+D^{*0}\bar{D}^0)$ and the same
element $\langle X|{\bf \sigma_2}\cdot{\bf\sigma_4}|X\rangle=+1$. In
the following calculation, we adopt the latter convention which is
consistent with the PDG assignment. So the quantum numbers for the
neutral states are $J^{PC}=1^{+\pm}$ corresponding to $P\bar{V}\pm
V\bar{P}$.

From the flavor SU(3) symmetry, it is easy to get the wave functions
of other systems in the same multiplet. One may use $P\bar{V}+c
V\bar{P}$ to denote these wave functions where $c$ is equivalent to
the C-parity of the neutral state.

\begin{table*}[htb]
\begin{tabular}{c|c||c|c}\hline
Isospin & $P\bar{V}+c V\bar{P}$ & $c=+1$ & $c=-1$ \\
\hline $I=\frac12$ & $\bar{D^0}D_s^{*+}+c  \bar{D}^{*0}D_s^+$ &
$\bar{D}^{*0}D_s^{*+}$ (J=2) & $\bar{D}^{*0}D_s^{*+}$ (J=1)\\
& $B^+\bar{B}_s^{*0}+c  B^{*+}\bar{B}_s^0$ & $B^{*+}\bar{B}_s^{*0}$
(J=2) & $B^{*+}\bar{B}_s^{*0}$ (J=1) \\
$I=1$ & $\bar{D^0}D^{*+}+c  \bar{D}^{*0}D^+$ &
$\bar{D}^{*0}D^{*+}$ (J=2) & $\bar{D}^{*0}D^{*+}$ (J=1) \\
& $B^+\bar{B}^{*0}+c  B^{*+}\bar{B}^0$ & $B^{*+}\bar{B}^{*0}$ (J=2)
& $B^{*+}\bar{B}^{*0}$ (J=1)\\
$I=0$ & $D_s^-D_s^{*+}+c  D_s^{*-}D_s^{+}$ & $D_s^{*-}D_s^{*+}$ (J=2) & $D_s^{*-}D_s^{*+}$ (J=1) \\
& $B_s^0\bar{B}_s^{*0}+c  B_s^{*0}\bar{B}_s^0$ &
$B_s^{*0}\bar{B}_s^{*0}$ (J=2) & $B_s^{*0}\bar{B}_s^{*0}$ (J=1) \\
& ($\bar{D}^0D^{*0}+c  \bar{D}^{*0}D^0$)+($D^-D^{*+}+c  D^{*-}D^+$) & $\bar{D}^{*0}D^{*0}+D^{*-}D^{*+}$ (J=2)& $\bar{D}^{*0}D^{*0}+D^{*-}D^{*+}$ (J=1) \\
& ($B^+B^{*-}+c  B^{*+}B^-$)+($B^0\bar{B}^{*0}+c  B^{*0}\bar{B}^0$) & $B^{*0}\bar{B}^{*0}+B^{*+}B^{*-}$ (J=2)& $B^{*0}\bar{B}^{*0}+B^{*+}B^{*-}$ (J=1) \\
\hline
$I=\frac12$ & $B^{*+}D_s^+$/$B^+D_s^{*+}$ & \multicolumn{2}{c}{$B^+D_s^+$}\\
$I=1$& $B^{*+}D^+$/$B^+D^{*+}$ & \multicolumn{2}{c}{$B^+D^+$} \\
$I=0$& $B_s^{*0}D_s^+$/$B_s^0D_s^{*+}$ & \multicolumn{2}{c}{$B_s^0D_s^+$} \\
& ($B^{*0}D^++B^{*+}D^0$)/($B^0 D^{*+}+B^+D^{*0}$) & \multicolumn{2}{c}{$B^0D^++B^+D^0$} \\
 \hline
ISB & $\bar{D}^{*0}D_s^+$ & \multicolumn{2}{c}{$\bar{D}^0D_s^+$ (I=$\frac12$)} \\
& $B^{*+}\bar{B}_s^0$ & \multicolumn{2}{c}{$B^+\bar{B}_s^0$ (I=$\frac12$)}\\
& $\bar{D}^{*0}D^+$ & \multicolumn{2}{c}{$\bar{D}^0D^+$ (I=1)} \\
& $B^{*+}\bar{B}^0$ & \multicolumn{2}{c}{$B^+\bar{B}^0$ (I=1)} \\
& $B^{*+}D^0$/$B^+D^{*0}$ & \multicolumn{2}{c}{$B^+D^0$ (ISB)}\\
& $\bar{D}^0D^{*0}+c  \bar{D}^{*0}D^0$ & $\bar{D}^{*0}D^{*0}$ (ISB) (J=2) & $\bar{D}^{*0}D^{*0}$ (ISB) (J=1) \\
& $B^+B^{*-}+c  B^{*+}B^-$ & $B^{*+}B^{*-}$ (ISB) (J=2) & $B^{*+}B^{*-}$ (ISB) (J=1)  \\
 \hline
\end{tabular}
\caption{The correspondence for the numerical results between
$P\bar{V}+c  V\bar{P}$ and $V\bar{V}$ or $P\bar{P}$. Here $c$ means
the C-parity of the neutral state of the multiplet.}\label{cores}
\end{table*}

In this pseudoscalar-vector case, the numerical results may be found
in the $P\bar{P}$ systems or the $V\bar{V}$ systems. We explain this
fact with I=1/2 states.

We investigate $\frac{1}{\sqrt2}(\bar{D^0}D_s^{*+}\pm
\bar{D}^{*0}D_s^+)$ and $\frac{1}{\sqrt2}(B^+\bar{B}_s^{*0}\pm
B^{*+}\bar{B}_s^0)$. By comparing the binding energies with I=1/2
$V\bar{V}$ case, one finds the results for the $c=+1$ ($c=-1$)
states are the same as those for $J=2$ ($J=1$)
$\bar{D}^{*0}D_s^{*+}$ or $B^{*+}\bar{B}_s^{*0}$. Therefore these
systems are also unbound.

It is unnecessary to consider $\frac{1}{\sqrt2}(B^+D_s^{*+}\pm
B^{*+}D_s^+)$ since the mass difference between $B^+D_s^{*+}$ and
$B^{*+}D_s^+$ is around 100 MeV and their mixing should be very
small. For the system $B^+D_s^{*+}$ or $B^{*+}D_s^+$, the results
are the same as $B^+D_s^+$ of the $P\bar{P}$ case.

Similarly, for the I=1 case, the results for the $c=+1$ ($c=-1$)
$\bar{D^0}D^{*+}$ and $B^+\bar{B}^{*0}$ are the same as $J=2$
($J=1$) $\bar{D}^{*0}D^+$ and $B^{*+}\bar{B}^0$, respectively. The
results for $B^+D^{*+}$ or $B^{*+}D^+$ are the same as $B^+D^+$
case. One can also get the results for I=0 cases and large ISB cases
from $J=2$ ($J=1$) $V\bar{V}$ or $P\bar{P}$. The correspondence for
the numerical results between $P\bar{V}\pm V\bar{P}$ and $V\bar{V}$
or $P\bar{P}$ may be found in Table \ref{cores}

Such a feature is not difficult to understand. The difference
between the $V\bar{V}$ case and the $P\bar{V}\pm V\bar{P}$ case
comes from the spin-spin parts of the potentials. The matrix element
for the $P\bar{V}\pm V\bar{P}$ is
$\langle\bm{\sigma}\cdot\bm{\sigma}\rangle=\pm1$ while that for the
$V\bar{V}$ is $\langle\bm{\sigma}\cdot\bm{\sigma}\rangle=-2$, $-1$
and $+1$ corresponding to $J=0$, $J=1$ and $J=2$, respectively.
Therefore the results for the $c=+1$ ($c=-1$) $P\bar{V}$ case are
similar to those for the $J=2$ ($J=1$) $V\bar{V}$ case. If
pseudoscalar meson exchanges are forbidden, the results for the
$P\bar{V}\pm V\bar{P}$ systems are similar to those for the
$P\bar{P}$.

\section{Discussions and conclusion}\label{sec6}

From the numerical results in the previous section, we know the
binding energy is always larger in the extended chiral quark model
than in the chiral quark model. This is partly because the vector
mesons provide attractive and relatively important interactions.
Another reason is that the sigma mass in E$\chi$QM is smaller than
that in $\chi$QM. This makes the attraction from $\sigma$ stronger
and thus the binding energy is larger even if the contributions from
vector mesons can be canceled.

In order to make a clearer picture for the possible hadronic
molecules, we summarize our conclusions in Tables \ref{sumppbar},
\ref{sumvvbar} and \ref{sumpvbar}. In those tables, ``$\times$''
means that a bound state does not exist. ``$\ast$'' means a bound
state does not exist in $\chi$QM while it is possible or not
excluded in E$\chi$QM. ``?'' means we cannot draw a conclusion even
in $\chi$QM and the system needs further study. ``$\checkmark$''
means a bound state is possible.

\begin{table}[htb]
\centering
\begin{tabular}{cccc}\hline
Isospin & ($\bar{c},c$) & ($\bar{b},b$) & ($\bar{b},c$) \\
$I=\frac12$ & $\times$ & $\times$ & $\times$ \\
$I=1$ & $\times$ & $\ast$ & $\times$ \\
$I=0(s)$ & $\ast$ & ? & ? \\
$I=0(l)$ & ? & $\checkmark$ & $\checkmark$
\\
ISB & $\ast$ & $\checkmark$ & ? \\ \hline
\end{tabular}
\caption{Summary of possible bound states in $P\bar{P}$ systems.
}\label{sumppbar}
\end{table}

\begin{table}[htb]
\centering
\begin{tabular}{cccc|ccc|ccc}\hline
 & \multicolumn{3}{c}{($\bar{c},c$)} & \multicolumn{3}{c}{($\bar{b},b$)} & \multicolumn{3}{c}{($\bar{b},c$)} \\
Isospin& J=0 & J=1 & J=2 & J=0 & J=1 & J=2 & J=0 & J=1 & J=2\\
$I=\frac12$ & $\times$ & $\times$ & $\times$ & $\times$ & $\times$ & $\times$ & $\times$ & $\times$ & $\times$  \\
$I=1$ & $\times$ & $\times$ & $\times$ & ? & ? & $\times$ & $\times$ & $\times$ & $\times$  \\
$I=0(s)$ & $\ast$ & $\ast$ & $\ast$ & ? & ? & ? & ? & ? & ? \\
$I=0(l)$ & $\ast$ & $\ast$ & $\checkmark$ & $\ast$ & ? & \checkmark
& $\ast$ & ? &
\checkmark\\
ISB & $\ast$ & $\ast$ & ? & $\ast$ & ? & $\checkmark$ & $\ast$ &
$\ast$ & ?
\\ \hline
\end{tabular}
\caption{Summary of possible bound states in $V\bar{V}$
systems.}\label{sumvvbar}
\end{table}

\begin{table}[htb]
\centering
\begin{tabular}{ccc|cc}\hline
 & \multicolumn{2}{c}{$C=+$}  & \multicolumn{2}{c}{$C=-$} \\
Isospin& ($\bar{c},c$) & ($\bar{b},b$) & ($\bar{c},c$) & ($\bar{b},b$)\\
$I=\frac12$ & $\times$ & $\times$ & $\times$ & $\times$ \\
$I=1$ & $\times$ & $\times$ & $\times$ & ?  \\
$I=0(s)$ & $\ast$ & ? & $\ast$ & ? \\
$I=0(l)$ & $\checkmark$ & \checkmark & $\ast$ & ? \\
ISB & $\ast$ & $\checkmark$ & $\ast$ & ? \\
\hline
\end{tabular}
\caption{Summary of possible bound states in $P\bar{V}\pm V\bar{P}$
systems.}\label{sumpvbar}
\end{table}

From the tables, we know that the I=1/2, I=1 charm-anticharm and the
I=1 bottom-charm hadronic molecules do not exist. Therefore our
conclusion for the $D^{*-}D_s^+\pm D^-D_s^{*+}$ system is
inconsistent with Ref. \cite{lee-mole}. Our calculation also
indicates that the resonance-like structure $Z^+(4051)$ in the
$\pi^+\chi_{c1}$ invariant mass \cite{twocharged} could not be an
S-wave $D^*\bar{D}^*$ molecule.

On the other hand, the isoscalar hidden bottom molecules $B\bar{B}$,
J=2 $B^*\bar{B}^*$, and C=+ $B\bar{B}^*$ are very likely to form
regardless of whether the isospin symmetry is largely violated or
not. All these states should be rather stable since $B$ is the
lowest bottom meson and $B^*$ do not decay via strong interaction.
The experimental search for these states may be used to test our
model.

There are so many systems we cannot draw a conclusion, most of which
are I=0 states. Whether the effects due to coupled channels, the
annihilation and the possible mixing between S-wave and D-wave
interactions may help is an open question. More detailed studies are
necessary.

In the extended chiral SU(3) quark model, the gluon, pseudoscalar,
scalar and vector mesons bind together the light quarks to baryons.
The strength of OGE interaction in the SU(3) chiral quark model is
greatly reduced due to the existence of vector mesons. But it is
still controversial whether OGE or vector meson exchange dominates
the short range quark-quark interaction. If the later mechanism is
not suitable, one finds more systems without binding solutions.

In summary, we have performed a systematic study for the bound state
problem of S-wave heavy quark meson-antimeson systems in a chiral
quark model. The exchanged mesons below 1.1 GeV have all been taken
into account. Since we considered just color-singlet meson-meson
configuration and several approximations were used, our
investigation is preliminary. Our crude calculation disfavors the
existence of I=1/2, I=1 charm-anticharm and I=1 charm-bottom
hadronic molecules but favors the existence of I=0 $B\bar{B}$,
$B^*\bar{B}^*$ (J=2) and $B\bar{B}^*$ (C=+) bound states. Whether
the consideration of other effects, such as the coupling with
hidden-color configuration and the coupling with possible D-wave,
supports these conclusions or not will be further studied. In our
model, the sigma meson exchange interaction plays an important role
in the bound state problem of the light quark systems. When
extending the model to the heavy quark sector, the possibility of
the sigma meson exchange between heavy quarks or between a heavy
quark and a light quark is not excluded. Since no mass factor in the
potential may suppress the sigma meson contributions, the value of
the coupling constant $g_{QQ\sigma}$ is crucial in discussing
whether or not such interactions are important. Although the
coupling is expected to be weak, a small value may have big effects,
which is also an open problem in the present approach.

\section*{Acknowledgments}

YRL thanks Professor S.L. Zhu, Professor B.S. Zou, Professor Q. Zhao
and Professor P.N. Shen for helpful discussions. This project was
supported by the National Natural Science Foundation of China under
Grants 10775146, 10805048, Ministry of Science and Technology of
China (2009CB825200), the China Postdoctoral Science foundation
(20070420526), and K.C. Wong Education Foundation, Hong Kong.

\end{document}